\documentclass[twocolumn]{aastex62}
\usepackage{amssymb}
\usepackage{amsmath}
\usepackage{array,multirow}
\usepackage{comment}
\usepackage{enumerate}
\usepackage{bm}

\newcommand{\Msun}{\ifmmode {M_{\odot}}\else${M_{\odot}}$\fi}
\newcommand{\Rsun}{\ifmmode {R_{\odot}}\else${R_{\odot}}$\fi}
\newcommand{\Porb}{\ifmmode {P_{\rm orb}}\else${P_{\rm orb}}$\fi}

\submitjournal{ApJL}

\shorttitle{DNS Mass Ratios}
\shortauthors{Andrews, Jeff J.}

\begin{document}

\title{{\bf Mass Ratios of Merging Double Neutron Stars as Implied by the Milky Way Population}}

\author[0000-0001-5261-3923]{Jeff J.\ Andrews}
\affiliation{Center for Interdisciplinary Exploration and Research in Astrophysics (CIERA), 
1800 Sherman Ave., 
Evanston, IL, 60201, USA}
\affiliation{Department of Physics and Astronomy, 
Northwestern University,
2145 Sheridan Rd., 
Evanston, IL 60208, USA}
\email{jeffrey.andrews@northwestern.edu}

\begin{abstract}
Of the seven known double neutron stars (DNS) with precisely measure masses in the Milky Way that will merge within a Hubble time, all but one has a mass ratio, $q$, close to unity. Recently, precise measurements of three post-Keplerian parameters in the DNS J1913$+$1102 constrain this system to have a significantly non-unity mass ratio of 0.78$\pm$0.03. One may be tempted to conclude that approximately one out of seven (14\%) DNS mergers detected by gravitational wave observatories will have mass ratios significantly different from unity. However J1913$+$1102 has a relatively long merger time of 470 Myr. We show that when merger times and observational biases are taken into account, the population of Galactic DNSs imply that $\simeq98\%$ of all merging DNSs will have $q>$0.9. We then apply two separate fitting formulas informed by 3D hydrodynamic simulations of DNS mergers to our results on Galactic DNS masses, finding that either $\simeq$0.004 \Msun\ or $\simeq$0.010 \Msun\ of material will be ejected at merger, depending on which formula is used. These ejecta masses have implications for both the peak bolometric luminosities of electromagnetic counterparts (which we find to be $\sim$10$^{41}$ erg s$^{-1}$) as well as the $r$-process enrichment of the Milky Way. 
\end{abstract}


\section{Introduction}
\label{sec:intro}

The characteristics of those merging double neutron stars (DNS) that are detected by the LIGO/Virgo gravitational wave observatories \citep{2017PhRvL.119p1101A, 2020ApJ...892L...3A} are the result of a combination of non-linear processes such as mass transfer and core collapse \citep[for a review, see][]{2017ApJ...846..170T}. The masses of DNSs are particularly sensitive to these physical processes. As such they have been the subject of many studies since \citet{1975ApJ...195L..51H} discovered the first system, and it was realized that post-Keplerian parameters \citep[e.g., Shapiro delay;][]{1964PhRvL..13..789S} could be used to measure the component masses \citep[for a review, see][]{2016ARA&A..54..401O}. One can use binary population synthesis to simulate the masses and mass ratios, $q$, of the subset of DNSs that will merge due to gravitational wave emission \citep[e.g.,][]{2011MNRAS.413..461O, 2012ApJ...759...52D, 2018MNRAS.481.4009V, 2018MNRAS.474.2937C, 2020arXiv200208011K}. However, much of the physics involved in DNS formation, in particular the masses, still lack complete, satisfactory descriptions. 

Alternatively, one can extrapolate the sample of known DNSs in the Milky Way as an indication of the mass ratios of merging DNSs in the local Universe. Initial models used maximum likelihood or Bayesian methods to fit the NS mass distribution with Gaussian distributions \citep{1999ApJ...512..288T, 2011MNRAS.414.1427V, 2012ApJ...757...55O, 2013ApJ...778...66K}. In the past five years, several studies have built upon these earlier works, adding sophistication and leveraging a steadily expanding observational data set \citep{2016arXiv160501665A, 2018MNRAS.478.1377A, 2018arXiv180403101H, 2019MNRAS.485.1665K, 2019ApJ...876...18F, 2019MNRAS.488.5020Z, 2020RNAAS...4...65F}. 

All these studies use the entire available population of DNSs with mass measurements to place their constraints. However, if one wants to extrapolate to the sample of merging DNSs, this approach is problematic for two reasons. First, as previously argued by \citet{2019ApJ...880L...8A}, the population of Galactic DNSs in the field likely evolved from at least three separate formation scenarios, one of which will not merge within a Hubble time. Since they form through different evolutionary pathways, these subpopulations are likely to have different underlying mass distributions. If one is interested in deriving mass constraints for LIGO/Virgo sources, the systems that do not merge in a Hubble time ought to be excluded from any analysis.

Second, not all systems ought to be weighted equally. Of the seven Galactic DNSs with well-measured masses that will merge in a Hubble time, only one system, J1913$+$1102 \citep[hereafter J1913;][]{2016ApJ...831..150L, 2018IAUS..337..146F} has $q<0.9$. Recently, \citet{2020arXiv200704175F} precisely measure the masses of the NSs in J1913 to be 1.62$\pm$0.03 and 1.27$\pm0.03$ \Msun, leading to $q=0.78\pm0.03$. These authors argue the existence of J1913 implies that $\simeq$11\% of merging DNSs ought to have $q<0.9$. As we show below, this argument is flawed as it takes into account neither the differing merger times nor the observational biases associated with each Galactic DNS.

\begin{figure}
\begin{center}
\includegraphics[width=\columnwidth,angle=0]{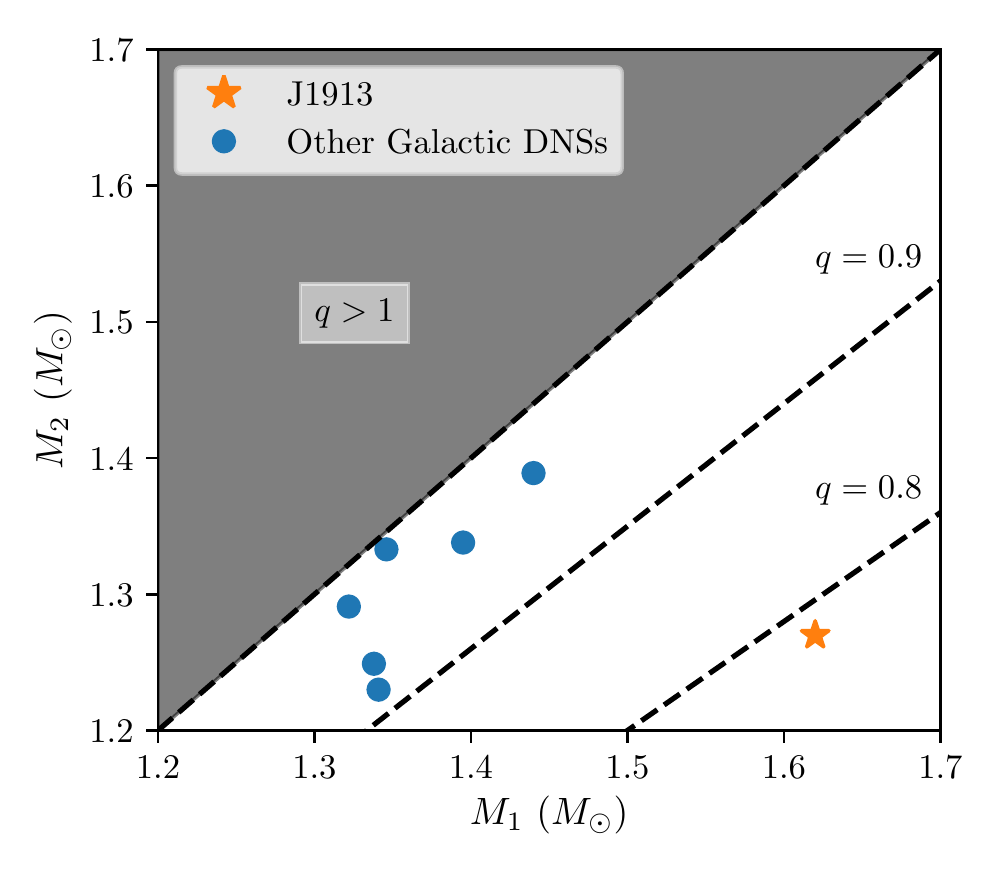}
\caption{ The mass distribution of DNS components for Galactic systems that will merge in a Hubble time. The mass of the more massive NS in the system is designated as $M_1$. J1913 is the only system of the seven with $q<$0.9. \label{fig:1}}
\end{center}
\end{figure}

This Letter is outlined as follows. In Section \ref{sec:method}, we describe our straightforward method, which is non-parametric, accounts for observational biases and merger times, and produces robust estimates under the assumption that the population of Galactic DNSs are representative of the population of DNSs in the local Universe. We then apply fitting formulas to these masses to calculate the distribution of ejecta masses at merger and associated electromagnetic counterpart luminosities in Section \ref{sec:ejecta}. Finally, we discuss our results and conclude in Section \ref{sec:conclusion}.

\section{Method}
\label{sec:method}

Of the 20 or so known Galactic DNSs, we focus on the seven systems in the field with precisely measured masses that will merge within a Hubble time\footnote{Throughout this work, we ignore systems associated with globular clusters, as these suffer from substantially different selection effects and are generally thought to comprise a minority of the overall population of Galactic DNSs \citep{2020ApJ...888L..10Y}.}. Listed in Table \ref{tab}, these include the six merging systems in \citet{2017ApJ...846..170T} as well as J1757$-$1854 \citep{2018MNRAS.475L..57C}. The recently detected J1946$+$2052 \citep{2018ApJ...854L..22S} still lacks precise mass measurements. In Figure \ref{fig:1}, we compare the NS masses in these systems; the difference between J1913 and its Galactic counterparts is glaringly apparent. However, to extrapolate from the $q$ distribution of the Galactic DNSs to the $q$ distribution of merging systems, two effects need to be accounted for: the observability through radio pulsar surveys and the merger time of each system.

\begin{table}[]
    \centering
      \begin{tabular}{lccccc}
  \hline
  System & $M_1$ & $M_2$ & $N_{\rm pop}$ & $\mathcal{R}$ & $\tau_{\rm merge}^a$ \\
   & ($M_{\odot}$) & ($M_{\odot}$) &  & (Myr$^{-1}$) & (Myr) \\
   \hline
  J0737$-$3039$^b$ & 1.338 & 1.249 & 1350$^{+780}_{-1350}$ & 5.8$^{+5.6}_{-3.4}$ & 86 \\
  B1534$+$12$^c$ & 1.346 & 1.333 & 1670$^{+1650}_{-970}$ & 0.6$^{+0.6}_{-0.3}$ & 2730 \\
  J1756$-$2251$^s$ & 1.341 & 1.230 & 1270$^{+1210}_{-740}$ & 0.8$^{+0.7}_{-0.4}$ & 1660 \\
  J1906$+$0746$^d$ & 1.322 & 1.291 & 690$^{+680}_{-400}$ & 11.3$^{+11.5}_{-6.4}$ & 309 \\
  J1913$+$1102$^f$ & 1.62 & 1.27 & 1560$^{+1530}_{-900}$ & 0.5$^{+0.5}_{-0.3}$ & 470 \\
  J1757$-$1854$^g$ & 1.395 & 1.338 & 1650$^{+970}_{-1590}$ & 10.0$^{+9.8}_{-5.8}$ & 76 \\
  B1913$+$16$^h$ & 1.440 & 1.389 & 2650$^{+2570}_{-1530}$ & 7.3$^{+7.1}_{-4.2}$ & 301 \\
  \hline
  \end{tabular}
  \caption{The list of Galactic DNSs with well-measured masses that will merge in a Hubble time. We provide the component masses, $M_1$ and $M_2$, where $M_1$ is always the more massive of the two. We ignore measurement uncertainties, as these are typically $\lesssim$0.01 $M_{\odot}$. References:
  $^a$\citet{2017ApJ...846..170T};
  $^b$\citet{2005tsra.conf..142K}; $^c$\citet{2014ApJ...787...82F}; $^d$\citet{2014MNRAS.443.2183F}; $^e$\citet{2015ApJ...798..118V}; $^f$\citet{2020arXiv200704175F}; $^g$\citet{2018MNRAS.475L..57C}; $^h$\citet{2010ApJ...722.1030W}.\\
   }
  \label{tab}
\end{table}

\begin{figure*}
\begin{center}
\includegraphics[width=\textwidth,angle=0]{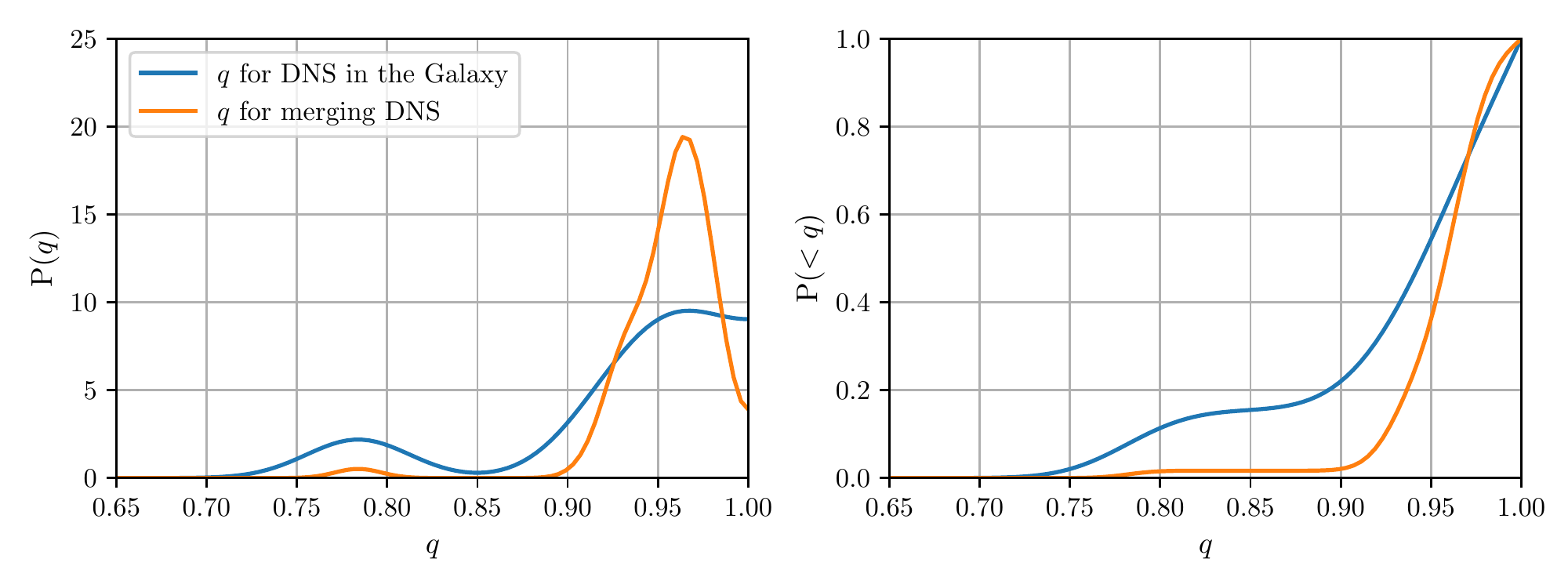}
\caption{ A KDE representation of the mass ratio of merging DNSs in the Milky Way with measured masses. The left panel shows the probability distribution for both the population expected to reside in the Milky Way at any one time (blue) as well as those DNSs that merge (orange). The right panel shows the corresponding cumulative distributions. If the population of Milky Way DNSs are representative of the sources that LIGO/Virgo detects, then $\simeq$98\% of systems have $q>0.9$. \label{fig:2}}
\end{center}
\end{figure*}

The observability of a DNS is affected by several factors, including the pulsar's luminosity, beaming factor, position in the Milky Way, Doppler smearing due to orbital motion, and the selection function of pulsar surveys. Depending on its observability, a single detected system may comprise the only one of its kind in the Milky Way, or it may represent the tip of an iceberg, implying a much larger, underlying population waiting to be identified with future, deeper pulsar surveys. Using the method pioneered by \citet{2003ApJ...584..985K}, one can quantify these effects for each system related to large-scale radio-pulsar surveys and calculate $N_{\rm pop}$, the number of DNSs in the Milky Way implied by each observed system. We refer to these as ``like" systems.

As discussed in detail by \citet{2003ApJ...584..985K}, the differing systems' merger times also affect our understanding of the underlying population in a similar way. To see this effect explicitly, consider two separate DNSs, one with a merger time of 10 Myr and one with 100 Myr. The system with a merger time of 10 Myr ought to be weighted 10 times its counterpart, since its detection implies that nine others have formed and merged during the lifetime of the longer-lived system. Note this bias only needs to be included when converting from $N_{\rm pop}$ into $\mathcal{R}$, the contribution to the overall Galactic DNS merger rate for each system. This bias is accounted for by weighting each system by the inverse of its merger time, $\tau_{\rm merge}$, listed in the last column of Table \ref{tab}.

We use the code provided by \citet{2019ApJ...870...71P, 2019ApJ...874..186P}\footnote{\href{https://github.com/devanshkv/PsrPopPy2}{https://github.com/devanshkv/PsrPopPy2} \citep[][Agarwal et al., in prep]{2014MNRAS.439.2893B}, \href{https://github.com/NihanPol/2018-DNS-merger-rate}{https://github.com/NihanPol/2018-DNS-merger-rate}}, which incorporates the latest descriptions of the pulsar surveys, to calculate $N_{\rm pop}$ and $\mathcal{R}$ for each system. Table \ref{tab} provides the median and 1-$\sigma$ confidence intervals for both of these parameters. Typical values of $N_{\rm pop}$ are 10$^3$ while $\mathcal{R} \sim 1$ Myr$^{-1}$.

We obtain numerical estimates of the relative rates of J1913-like systems in the Milky Way at any one time, $f_{\rm J1913, Milky Way}$, accounting for uncertainties using a Monte Carlo method:
\begin{equation}
    f_{\rm J1913,\ Milky\ Way} = \frac{\sum_{j=1}^{N}N_{\rm pop, J1913, j}}{\sum_{i\in {\rm sys}}\sum_{j=1}^{N}N_{\rm pop, i, j}} = 16\%.
\end{equation}
The summations over $j$ are to propagate the errors on $N_{\rm pop}$ using Monte Carlo sampling. So for each of the $i$ systems, we draw $N=100$ values of $N_{\rm pop, i, j}$ from the likelihood distribution we calculated using the code from \citet{2019ApJ...870...71P}. The summation in the denominator is a normalization factor to account for each of the $i$ systems in our sample. Our resulting prevalence of J1913-like systems in the Milky Way is 16\%, consistent with the rate of $11^{+21}_{-9}$\% derived by \citet{2020arXiv200704175F}. However, if we want to derive the relative contribution of J1913-like systems to the merger rate of Galactic DNSs, we need to instead weigh each system by $\mathcal{R}$:
\begin{equation}
    f_{\rm J1913,\ merging\ DNS} = \frac{\sum_{j=1}^{N}\mathcal{R}_{\rm J1913, j}}{\sum_{i\in {\rm sys}}{\sum_{j = 1}^{N}{\mathcal{R}_{\rm i, j}}}} = 2\%, \label{eq:frac}
\end{equation}
where $\mathcal{R}_{\rm i, j}$ are samples randomly drawn from the likelihood distribution over each of the $i$ systems. We find that since J1913 contributes so little to the overall merger rate, only $\simeq$2\% of merging DNSs in the Milky Way are expected to have such low $q$.

We represent the $q$ distribution of merging DNSs using a kernel density estimate (KDE)\footnote{To compute our KDEs, we use \texttt{kalepy}; \href{https://github.com/lzkelley/kalepy}{https://github.com/lzkelley/kalepy}. We use a Gaussian kernel, with a bandwidth of 0.4 and a reflecting boundary at $q=1$.} in Figure \ref{fig:2}, where individual points are weighted by either $N_{\rm pop}$ (blue) or by $\mathcal{R}$ (orange). The left panel shows the probability density function, while the right panel shows the cumulative distribution. When the distribution is weighed by $N_{\rm pop}$, which represents the distribution of DNSs that exist at any one time in the Milky Way, a non-trivial fraction of systems have $q<$0.9. However, when weighted by $\mathcal{R}$ so as to represent the distribution of merging DNSs, we find that 98\% of all merging DNSs have $q>0.9$, in agreement with Equation \ref{eq:frac}. While the exact details of the distribution are dependent upon specifics of how the KDE representation is computed, using any set of reasonable values our main conclusion that nearly all merging DNSs have $q>0.9$ is robust.

We re-emphasize that an extrapolation to the sample of LIGO/Virgo sources relies on the assumption that the Galactic systems are representative of the DNS population in the local Universe. While the method of \citet{2003ApJ...584..985K} attempts to account for observational biases, imperfections may persist. Furthermore, our conclusions are based on the mere seven known merging systems with measured masses. Because of the weighting scheme, the detection by pulsar surveys of even one new system with a sufficiently short merger time (and therefore, large $\mathcal{R}$) can significantly alter the derived $q$ distribution. Clearly, the radio detection of new DNSs in the Milky Way with properties substantially different from those already known will further refine the $q$ distribution of merging DNSs.

\begin{figure}
\begin{center}
\includegraphics[width=\columnwidth,angle=0]{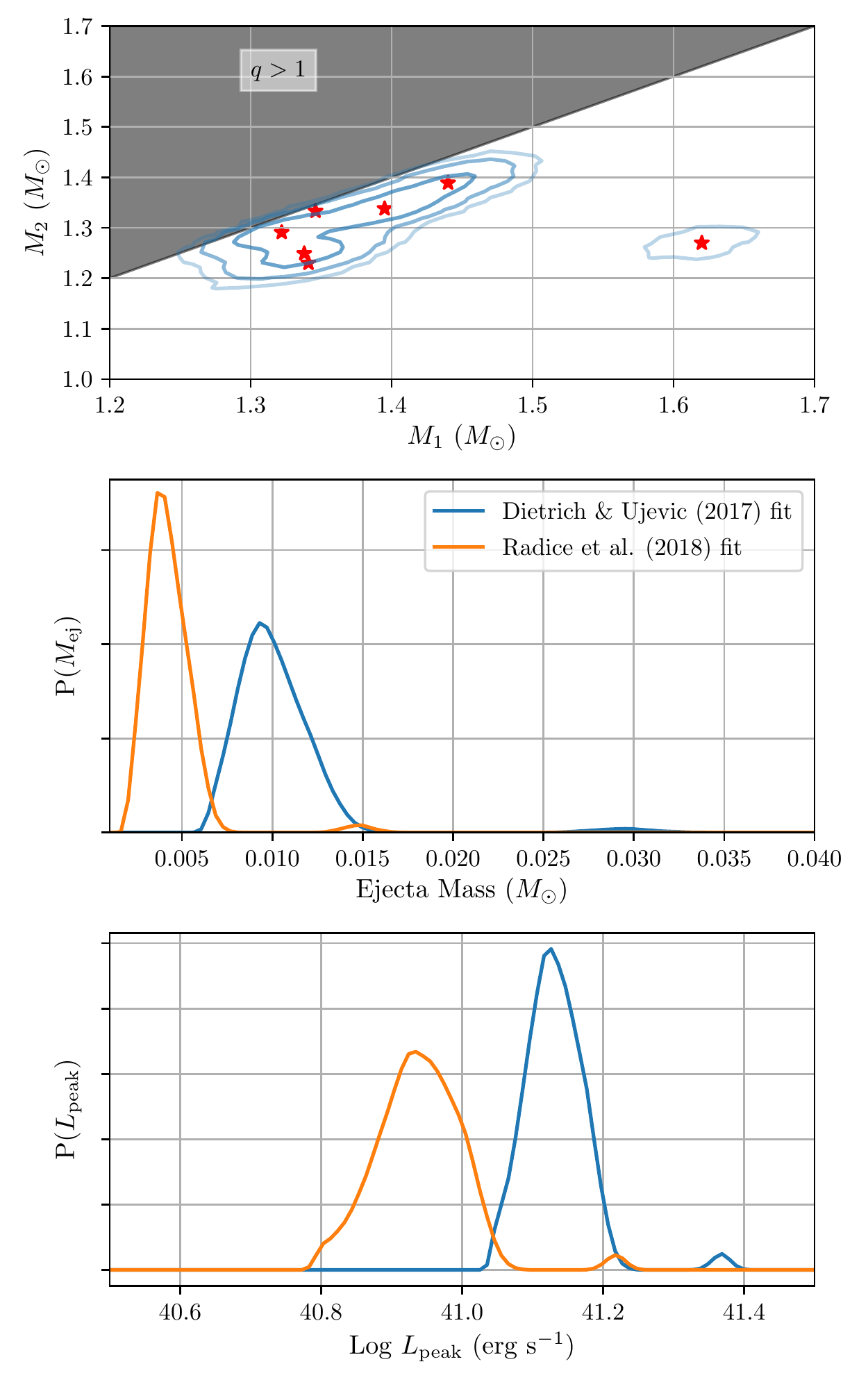}
\caption{ The top panel shows the 1-, 2-, and 3-$\sigma$ contours of the component masses of DNS mergers, taken from the KDE representation of the Galactic DNS population (red stars) weighted by their relative merger rates. In the middle panel, we calculate the distribution of ejecta masses from DNS mergers, derived from fitting formula, from both \citet{2017CQGra..34j5014D} and \citet{2018ApJ...869..130R}, applied to our KDE representations of merging DNS masses. Both fitting formulas show ejecta masses dominated by a single peak, with J1913 contributing a nearly insignificant higher mass second peak. From the ejecta masses, we use the `Arnett rule' to estimate the peak luminosity of an associated electromagnetic counterpart. Depending on the choice of parameters chosen, we find peak luminosities of $\sim10^{41}$ erg s$^{-1}$.  \label{fig:3}}
\end{center}
\end{figure}

\section{Ejecta Masses and Luminosities}
\label{sec:ejecta}

We can further use the population of Galactic DNSs to derive expectations for the ejecta masses of DNS mergers. We first use \texttt{kalepy} to produce a KDE representation\footnote{We again use a Gaussian kernel with a bandwidth of 0.4 and a reflecting boundary at $q=1$.} of the 2D $q-M_{\rm tot}$ distribution (previously we fit the 1D distribution of $q$). We show the resulting distribution, transformed into $M_1-M_2$ space and weighted by $\mathcal{R}$, as 1-, 2-, and 3-$\sigma$ contours in the top panel of Figure \ref{fig:3}. The outlier, J1913, only contributes at the 3-$\sigma$ level.

Using \texttt{kalepy} we obtain random variates of $M_1$ and $M_2$ drawn from this KDE representation. For each NS, we further calculate its baryonic mass $M^*$ using the formula $M_x^* = M_x + 0.075 M_x^2$ \citep{1989ApJ...340..426L}, and its compactness $C_x = \mathcal{G} M_x / c^2 R_x$. For simplicity, we assume all NSs have a radius, $R_x$, of 11 km. Using these parameters, we calculate the ejecta masses expected from a DNS merger employing the fitting formula from \citet{2017CQGra..34j5014D}:
\begin{equation}
\begin{split}
    \frac{M_{\rm ej}}{10^{-3} M_{\odot}} &= \left[ \alpha \left( \frac{M_a}{M_b} \right)^{1/3} \left( \frac{1-2C_a}{C_a} \right) + \beta \left( \frac{M_b}{M_a} \right)^n \right. \\
    &  \left. \quad + \gamma \left(1 - \frac{M_a}{M_a^*} \right) \right] M_a^* + (a \leftrightarrow b) + \delta.
\end{split}
\label{eq:3}
\end{equation}
By running a broad array of 3D hydrodynamic simulations of DNS mergers, where the NS masses are varied, and then fitting the coefficients, \citet{2017CQGra..34j5014D} finds values for the parameters of: $\alpha=-1.94315$, $\beta=14.9847$, $\gamma=-82.0025$, $\delta=4.75062$, $n=-0.87914$. \citet{2018ApJ...869..130R} generate a similar suite of simulations, finding values of $\alpha=-0.657$, $\beta=4.254$, $\gamma=-32.61$, $\delta=5.205$, $n=-0.773$. 

The middle panel of Figure \ref{fig:3} shows the resulting distribution of ejecta masses. The near-unity mass ratios imply relatively low ejecta masses of $\simeq 0.01$ and $\simeq 0.004$ \Msun\ for the fits from \citet{2017CQGra..34j5014D} and \citet{2018ApJ...869..130R}, respectively. Both distributions show a slight secondary peak with somewhat higher ejecta masses designating the contribution from J1913. Due to its relatively small $\mathcal{R}$ value, J1913's contribution to the ejecta mass distribution is at the 2\% level. 

Radioactive heating of this ejecta causes an electromagnetic transient \citep{1998ApJ...507L..59L} that can be observed with targeted follow-up of a gravitational wave event \citep{2017ApJ...848L..12A}. The `Arnett rule' \citep{1982ApJ...253..785A} provides an estimate of the dependence between ejecta mass and the corresponding bolometric luminosity \citep{2010MNRAS.406.2650M, 2014MNRAS.441.3444M, 2016AdAst2016E...8T}:
\begin{equation}
    \begin{split}
    L_{\rm peak} &\simeq 1.4 \times 10^{41}\ {\rm erg}\ {\rm s}^{-1} \left( \frac{f}{3\times 10^{-6}} \right) \left( \frac{v_r}{0.1 c} \right)^{1/2} \\
    & \quad \times \left( \frac{M_{\rm ej}}{0.01\ \Msun} \right)^{1/2} \left( \frac{\kappa}{10\ {\rm cm}\ {\rm g}^{-1}} \right)^{-1/2},
    \end{split}
    \label{eq:4}
\end{equation}
where $f$ quantifies the fraction of radioactive energy deposited in the material, $v_r$ is the expansion velocity, and $\kappa$ is the opacity. The bottom panel of Figure \ref{fig:3} shows the distribution of electromagnetic luminosities, calculated using Equation \ref{eq:4} and the distribution of ejecta masses displayed in the middle panel of Figure \ref{fig:3}. Due to the weak dependence on ejecta mass, the model in Equation \ref{eq:4} for the $L_{\rm peak}$ produces a distribution narrowly focused around 10$^{41}$ erg s$^{-1}$. However, one ought to take these results as only an order of magnitude estimate, since these luminosities are produced from a simplified description assuming spherical symmetry. For instance, they do not account for variations in $v_c$ and $\kappa$ as a function of viewing angle.

\section{Discussion and Conclusions}
\label{sec:conclusion}

Recently, \citet{2020arXiv200704175F} measure a mass ratio of 0.78 for the DNS J1913$+$1102. Since this is one of eight merging DNSs in the Milky Way, these authors argue that their observation implies that $\simeq$11\% of DNS mergers will have mass ratios significantly different from unity. Given its low mass ratio, \citet{2020arXiv200704175F} suggest that J1913 could be a Milky Way analog to the DNS merger forming GW170817, as its electromagnetic counterpart implies significant mass loss \citep{2017ApJ...848L..17C, 2017Natur.551...80K}, and therefore potentially a non-unity mass ratio. Interestingly, \citet{2019ApJ...883L...6R} also suggested that J1913$+$1102 was a Galactic analog to the progenitor of GW170817, albeit for entirely different reasons, based on the X-ray afterglow time delay. 

However the prevalence of low mass ratio systems calculated by \citet{2020arXiv200704175F} does not take into account either: a) observational biases associated with different pulsars and pulsar surveys or b) the different merging times of each system. Our analysis suggests that once you properly include these effects, nearly all ($\simeq98\%$) of the merging DNSs in the Milky Way have mass ratios larger than 0.9. We additionally apply fitting formula to our KDE representation of NS masses to determine the distribution of ejecta masses during the DNS merger. We find that typical masses ejected are $\simeq$0.004\Msun\ and 0.010\Msun\ using the fitting formulas from \citet{2018ApJ...869..130R} and \citet{2017CQGra..34j5014D}, respectively.  

We use the `Arnett rule' to estimate peak luminosities of the electromagnetic counterparts to our derived population of merging DNS \citep{2014MNRAS.441.3444M}. As an order of magnitude estimate, we find peak luminosities $\sim$10$^{41}$ erg s$^{-1}$. Alternatively, one could use these ejecta masses to compute a series of lightcurves corresponding the sample of DNS merger events detected by LIGO/Virgo \citep{2013ApJ...775...18B}. When combined with the sensitivities and field-of-view of a particular telescope, these synthetic lightcurves could be used to optimize a search strategy for a putative electromagnetic counterpart.

At the same time, the ejecta masses we calculate have implications for the $r$-process enrichment of the Milky Way \citep{1999A&A...341..499R, 2018IJMPD..2742005H}. For instance \citet{2018ApJ...860...89M} find that, whatever the origin of $r$-process material is, it must produce $\gtrsim$10$^{-3}$\Msun\ per event to explain the $r$-process abundances of halo stars in the Milky Way. This lower limit would fit with our distribution of ejecta masses shown in the middle panel of Figure \ref{fig:2}. However, there is disagreement in the literature, as some authors \citep[e.g.,][]{2019EPJA...55..203S} have argued that collapsar models may have less difficulty explaining the details of $r$-process observations.

Our analysis differs from recent previous studies that focus on the masses of NSs in DNS systems \citep[e.g.,][]{2019ApJ...876...18F} in three important ways. First, whereas previous studies fit all DNSs with mass measurements, we focus only on those DNSs in the field that merge within a Hubble time. This choice is motivated by our interest in the progenitors to gravitational wave merger events. Second, we weigh each system by its individual merger rate \citep[calculated using the method of][]{2003ApJ...584..985K}, such that systems with short merger times are more heavily weighted; for every system we observe, there are many more that have been formed and already merged. This quantitative analysis also takes into account observational biases associated with the detection of pulsar binaries. Third, rather than confining ourselves to a parametric model, which will suffer if the model chosen proves to be an inaccurate representation of the underlying distribution (e.g., trying to fit a non-Gaussian distribution with a Gaussian model), we use a non-parametric method to describe the mass distribution of merging DNSs in the Milky Way\footnote{Of course, our KDE approach has its own limitations, as it introduces hyperparameters associated with the shape and bandwidth of the kernel used. Nevertheless, our main conclusions are robust for any reasonable set of hyperparameters}. As a result of these three differences, our model produces more stringent limits. For example, using Gaussian models separately fit to the recycled pulsar and its companion \citet{2019ApJ...876...18F} find that 83.6\% of all DNSs have $q>0.9$ under their best-fit hypothesis (compared with our finding that $\simeq$98\% have $q>0.9$). 

Can these results be extrapolated to the local Universe to infer the properties of LIGO/Virgo detections? There are reasons to think not. \citet{2018ApJ...866...60P} suggests that GW170817, the first DNS merger detected by LIGO/Virgo has a mass ratio too low to be represented by the Galactic DNS population. Furthermore, the second LIGO/Virgo detection of a DNS merger, \citep[GW190425;][]{2020ApJ...892L...3A}, has a total mass significantly larger than any known system in the Milky Way. Yet, a resolution may be possible; the strong weighting by merger time implies that the detection of a single DNS with a short merger time can significantly alter the census of merging DNSs in the Milky Way. If future observations, both by radio telescopes and gravitational wave observatories cannot resolve this discrepancy, we may be forced to admit the possibility that the Milky Way DNS population forms a poor representation of the local Universe.

\acknowledgments

The author thanks Tassos Fragos, Vicky Kalogera and Christopher Berry for a careful reading of a previous version of the draft on short notice. The author also acknowledges funding from CIERA and Northwestern University through a Postdoctoral Fellowship.

\software{\texttt{kalepy}, \texttt{PsrPopPy2} \citep[][Agarwal et al., in prep]{2014MNRAS.439.2893B}, \texttt{2018-DNS-merger-rate} \citep{2019ApJ...870...71P}}

\bibliographystyle{aasjournal}

\end{document}